\documentclass[onecolumn,amsmath,amssymb,12pt,superscriptaddress,nofootinbib]{revtex4}
\pdfoutput=1

\usepackage[latin1]{inputenc}
\usepackage[english]{babel}
\usepackage{amssymb}
\usepackage{amsmath}
\usepackage{amsthm}
\usepackage[]{graphicx}
\usepackage[]{subfigure}
\usepackage{tensor}
\usepackage{color}
\usepackage{cancel}
\usepackage{setspace}
\usepackage{fancyhdr}
\usepackage[bookmarks,linktocpage, colorlinks=true, plainpages = false, citecolor = blue,  linkcolor=blue, urlcolor = blue, filecolor = blue]{hyperref}

\begin{document}

\allowdisplaybreaks
\begin{titlepage}

\title{Small-Field and Scale-Free: Inflation and Ekpyrosis at their Extremes \vspace{.3in}}

\author{Jean-Luc Lehners}
\email{jlehners@aei.mpg.de}
\affiliation{Max--Planck--Institute for Gravitational Physics (Albert--Einstein--Institute), 14476 Potsdam, Germany}

\begin{abstract}
\vspace{.2in} \noindent There is increasing evidence from string theory that effective field theories are only reliable over approximately sub-Planckian field excursions. The two most promising effective models for early universe cosmology, inflation and ekpyrosis, are mechanisms that, in order to address cosmological puzzles, must operate over vast expansion/energy ranges. This suggests that it might be appropriate to describe them using scaling laws. Here we combine these two ideas and demonstrate that they drive inflation and ekpyrosis  to their extremes:  inflation must start at ultra-slow-roll, and ekpyrosis at ultra-fast-roll. At face value, the implied spectra are overly tilted to the red, although in both cases minor departures from pure scale freedom bring the spectral indices within current observational bounds. These models predict a significant spectral running at a level detectable in the near future ($\alpha_s \approx -10^{-3}$). Ekpyrotic models with minimal coupling are nevertheless ruled out, as they lead to levels of non-Gaussianity that are at least an order of magnitude too large. No such restrictions apply to models with a kinetic coupling between the two ekpyrotic scalar fields, and these remain the most promising ekpyrotic models. 

An additional swampland criterion that was recently proposed for the slope of the scalar field potential would however rule out all ultra-slow-roll models of inflation. Finally, we speculate on the existence of corresponding restrictions on the slope at negative potential values, which might lead to similarly severe constraints on ekpyrotic models.

\end{abstract}
\maketitle

\end{titlepage}

\tableofcontents


\section{Introduction}

For a long time, cosmological model building rested on the premise that essentially any 4-dimensional effective field theory provided a justified starting point. But there is now increasing evidence that there may exist strong restrictions on what kind of effective theory is allowed in quantum gravity, or at the very least in string theory. This is a highly welcome development, as it implies that string theory finally puts restrictions on 4-dimensional, low-energy properties of the world. There is thus a shift of perspective from an ``anything-goes'' attitude to Vafa's swampland \cite{Vafa:2005ui,Brennan:2017rbf} with its surprising connections between gravity and particle physics for consistent low-energy theories. For example, the weak gravity conjecture has been used to derive a connection between the scale of dark energy and the Higgs and neutrino masses \cite{Ibanez:2017kvh}. For cosmology, two principles are especially relevant: the field excursion bound \cite{Ooguri:2006in}, and the newly proposed slope bound \cite{Obied:2018sgi,Agrawal:2018own,Andriot:2018wzk} -- see also \cite{Baume:2016psm,Klaewer:2016kiy,Danielsson:2018ztv,Blumenhagen:2018hsh,Garg:2018reu} for further studies of these bounds.

In the present paper, these principles will be applied to inflation and ekpyrosis, in combination with the hydrodynamical approach introduced by Mukhanov \cite{Mukhanov:2013tua} and developed by Ijjas, Steinhardt and Loeb \cite{Ijjas:2013sua}.  More specifically, we will take the attitude that since inflation has to work over an enormous range of scale factor values, and since ekpyrosis has to work over an equally enormous range of energy scales, it is natural to assume that they must operate according to scaling laws. Putting these physical principles together, i.e. imposing a small field range and adopting a description in terms of scaling laws, implies surprisingly strong constraints on early universe cosmological models. In the inflationary context, such an approach was already considered in \cite{Garcia-Bellido:2014wfa} while our results regarding ekpyrosis are new. For completeness the inflationary calculation will however also be reviewed below. In all cases we assume the best case scenario in terms of initial conditions, i.e. we will assume that the scalar field will start at rest and with a spatially uniform distribution at the desired location of the scalar field potential. The fundamental origin of such initial conditions remains elusive to date \cite{Feldbrugge:2017mbc}, and will ultimately require a convincing explanation, but this will not be the topic of the present work.

Our setting will be ordinary general relativity minimally coupled to a scalar field $\phi$ with a potential $V(\phi),$ so that the action is given by
\begin{align}
S = \frac{1}{2}\int d^4x \sqrt{-g} \left[ R - (\partial\phi)^2 - 2 V(\phi)\right]\,,
\end{align}
where we use reduced Planck units, $8\pi G=1.$ In a flat Robertson-Walker background $ds^2 = -dt^2 + a^2(t)d\underline{x}^2,$ and assuming the scalar depends only on time, the constraint and equations of motion are given by
\begin{align}
3H^2 &= \rho\,, \\
\dot{H} & = - \frac{1}{2}(\rho + p)\,, \\
\dot\rho + 3H (\rho + p) &= 0\,,
\end{align}
where the expansion rate is denoted by $H=\dot{a}/a,$ and where the energy density $\rho$ and the pressure $p$ are given by
\begin{align}
\rho = \frac{1}{2}\dot\phi^2 +  V(\phi)\,, \qquad & p = \frac{1}{2}\dot\phi^2 -  V(\phi)\,.
\end{align}
A highly useful quantity is the slow-roll/fast-roll parameter
\begin{align}
\epsilon = \frac{3}{2}\left( 1+\frac{p}{\rho}\right)\,, \label{epsilon}
\end{align}
which has a simple dependence on the equation of state $p/\rho.$ Inflation occurs when $H>0$ and $\epsilon < 1,$ since these conditions are equivalent to demanding accelerated expansion. By contrast, an ekpyrotic phase is defined as corresponding to contraction ($H<0$) with $\epsilon > 3,$ with the latter condition ensuring that the ekpyrotic field's energy density ($\propto a^{-2\epsilon}$) grows faster than that contained in small anisotropies of the metric ($\propto a^{-6}$). We will discuss both theories in turn, starting with a review of results on inflation \cite{Mukhanov:2013tua,Ijjas:2013sua,Garcia-Bellido:2014wfa,Boubekeur:2014xva}.

\section{Inflation}

The duration of inflation can be characterised by looking at the number of e-folds $N$ of expansion that occur before the end of inflation,
\begin{align}
dN \equiv - d\, \ln(a)\,,
\end{align}
where the end of inflation is reached when the slow-roll parameter $\epsilon$ reaches the value $1.$ As a consequence, a scale-free description of inflation is achieved by considering the scaling law \cite{Ijjas:2013sua}
\begin{align}
\epsilon^i = \frac{1}{(N+1)^{\gamma^i}}\,,
\end{align}
where $\gamma^i$ is a parameter and the superscript $i$ will be reserved for inflationary quantities. In realistic models, $N$ would then run from a value of at least about $60$ (depending on the details of reheating) down to $0,$ where the inflationary phase would end. In order to derive the associated field range, one may make use of the many re-writings of the slow-roll/fast-roll parameter defined in Eq. \eqref{epsilon}, which can equivalently be defined or expressed as
\begin{align}
\epsilon \equiv \frac{d\, \ln(H)}{dN}= - \frac{\dot{H}}{H^2} = \frac{1}{2}\frac{d\, \ln(\rho)}{dN}\,.
\end{align}
These relations allow one to find a relation between the field and the number of e-folds, 
\begin{align}
\frac{d\phi}{dN} = - \sqrt{2\epsilon}\,,
\end{align}
which one can integrate to find the inflationary field range $\Delta\phi^i$ (when $\gamma \neq 2$)
\begin{align}
\Delta\phi^i = \frac{2\sqrt{2}}{\gamma - 2}\left[ (\Delta N+1)^{\frac{2-\gamma}{\gamma}} - 1\right]\,. \label{DeltaPhiInf}
\end{align}
Note that had we treated the slow-roll parameter as constant (which we refer to as the ``naive'' case, since it does not take into account the fact that inflation must come to an end), this expression would have simplified to 
\begin{align}
\Delta\phi^i_{naive} = \sqrt{2\epsilon} \Delta N\,. \label{DeltaPhiInfNaive}
\end{align}
We are now in a position to determine the consequences of imposing a bound on the total field excursion -- in particular we will impose that the field range should be at most $\Delta\phi^i = 1$ in reduced Planck units. The expressions for the field ranges \eqref{DeltaPhiInf} and \eqref{DeltaPhiInfNaive} then imply that the parameter $\gamma^i$ must be larger than a lower bound that we call $\gamma^i_{min}$ and which for the two cases of constant and scaling $\epsilon$ is given by
\begin{align}
\gamma^i_{min,naive} = 2.1 \,&\rightarrow \,\epsilon^i=1.8 \times 10^{-4}\,, \\
\gamma^i_{min,scaling} = 4.8 \,&\rightarrow \,\epsilon^i_{60} = 2.7 \times 10^{-9}\,.
\end{align}
We can see that for constant $\epsilon$ the required value is already very small. However, in the physically more appealing case of scale-free $\epsilon,$ inflation is required to start out at ultra-slow-roll values of $\epsilon^i_{60} \lessapprox 10^{-9}.$ In other words, the implied scalar potential would start out being ultra-flat, and it would gradually steepen until the end of inflation is reached. Here the notation $\epsilon^i_{60}$ refers to the value of the slow-roll parameter at $N=60,$ i.e. at about the time when the currently observable fluctuations are amplified. This brings us to the topic of fluctuations and observational consequences of these models. Standard calculations \cite{Mukhanov:1981xt,Bardeen:1983qw,Kosowsky:1995aa,Starobinsky:1979ty} show that the spectral tilt $n_s$ of the scalar density fluctuations, its running $\alpha_s$ (i.e. the change of the spectral index with frequency $k,$ more explicitly $\alpha_s = \frac{d\,n_s}{d\,\ln(k)} \approx - \frac{d\,n_s}{dN}$ at Hubble crossing $k = aH$) and the tensor-to-scalar ratio $r$ are given by
\begin{align}
n_s  & = 1 - \frac{2}{(N+1)^\gamma} - \frac{\gamma}{N+1}\,, \\
\alpha_s &  = - \frac{2\gamma}{(N+1)^{\gamma+1}} - \frac{\gamma}{(N+1)^2}\,, \\
r &= 16 \epsilon\,.
\end{align}
In the present case we then obtain the predictions
\begin{align}
n^i_{s,naive} = 0.97 & \qquad n^i_{s,scaling} = 0.92 \,,\\
\alpha^i_{s,naive} = -5.8 \times 10^{-4} & \qquad \alpha^i_{s,scaling} = -1.3 \times 10^{-3} \,,\\
r^i_{naive} = 0.003 & \qquad r^i_{scaling} = 4.3 \times 10^{-8}\,.
\end{align}
Note that the naive tilt is spot-on regarding current observational bounds (as we will discuss in more detail below). Also, the running is small, which is unsurprising since we have a constant equation of state, and the level at which gravitational waves would be expected is again small, but within reach of near-future experiments. But such models with constant equation of state fail to take into account that inflation must come to an end. If we want this evolution to be smooth, rather than abrupt, the scaling model is a natural candidate to look at. Interestingly, it predicts a tilt that is very red. Moreover, this model leads to a significant running, which again is intuitively clear as the equation of state is changing. Finally, the fact that at 60 e-folds from the end the equation of state must be absolutely tiny implies that the strength of long-wavelength primordial gravitational waves is truly tiny, at a level that is essentially undetectable. Below we will discuss the comparison to observations in more detail -- for now, the main lesson is that the combination of demanding a small field range and a scaling law imposes highly non-trivial constraints on inflationary models, with interesting consequences regarding observational signatures. We can now compare this with ekpyrosis.

\section{Ekpyrosis}

In ekpyrosis, the standard cosmological puzzles (regarding flatness, causality and inhomogeneities) are addressed during a contracting phase taking place prior to the current expanding phase \cite{Khoury:2001wf,Lehners:2008vx}. In these models, the change from contraction to expansion is typically envisioned to occur via a cosmological bounce, for which many models are being explored \cite{Khoury:2001bz,Lehners:2006pu,Buchbinder:2007ad,Creminelli:2007aq,Easson:2011zy,Gielen:2015uaa,Ijjas:2016vtq,Farnsworth:2017wzr,Bramberger:2017cgf}. For us, the most important aspect of these bounces is that they occur on small (microphysical) scales, and that we may thus expect that they will not significantly alter the properties of long-wavelength perturbations seen in the CMB sky. Concrete calculations support this point of view, see e.g.  \cite{Battarra:2014tga}. But first we should discuss the background evolution. During an ekpyrotic phase the universe slowly contracts according to $a(t) \sim (-t)^{1/\epsilon},$ with $\epsilon > 3.$ In contrast to inflation, the requirement on $\epsilon$ does not come from demanding a solution to the flatness problem, as in a contracting universe the relative energy density of homogeneous curvature (which scales as $1/a^2$) naturally becomes subdominant in any case. In other words, a contracting universe does not have a flatness problem in the usual sense. However, small anisotropies in the metric (which in an expanding universe quickly decay away) grow rather fast during a contraction phase, as they scale as $a^{-6}.$ In order for the universe not to become dominated by anisotropies, which would entail a complete loss of predictivity, one demands that the ekpyrotic matter (scaling as $a^{-2\epsilon}$) becomes more dominant than anisotropies, i.e. one demands that $\epsilon > 3.$ The length of such an ekpyrotic phase is then usefully characterised by the change in $aH,$ and we will define the number of e-folds ${\cal N}$ of ekpyrosis left before the end via the relation
\begin{align}
 d{\cal N} \equiv - d\, \ln(aH)\,.
\end{align}
Thus the change in $aH$ is related to the change in the scale factor alone via
\begin{align}
d{\cal N} = (\epsilon - 1)dN\,. \label{NtoN}
\end{align}
Ekpyrosis is a mechanism that must operate over a vast range of energy scales in order to enhance flatness/isotropy and to produce cosmological perturbations on the relevant scales, and in analogy with inflation we will consider the equation of state (fast-roll parameter) to evolve according to the power law \cite{Ijjas:2013sua}
\begin{align}
\epsilon^e = 3 ({\cal N}+1)^{\gamma^e}\,,
\end{align}
which ensures that the ekpyrotic phase comes to an end when ${\cal N}$ reaches zero. The superscript $e$ will be reserved for ekpyrotic quantities. Using Eq. \eqref{NtoN}, we can then find an expression for the infinitesimal field range,
\begin{align}
\frac{d\phi}{d{\cal N}} = - \frac{\sqrt{2\epsilon}}{\epsilon - 1}\,,
\end{align}
which can be integrated to yield the total ekpyrotic field excursion
\begin{align}
\Delta\phi^e = - \sqrt{6}\int_{\Delta{\cal N}}^0 \frac{({\cal N}+1)^{\gamma/2}}{3({\cal N} +1)^\gamma-1} d{\cal N}\,.
\end{align}
If we had assumed a constant equation of state, which we again refer to as the ``naive'' model, we would have obtained the simpler expression
\begin{align}
\Delta\phi^e_{naive} = \frac{\sqrt{2\epsilon}}{\epsilon - 1}\Delta {\cal N}\,.
\end{align}
We can now explore the consequences of imposing the field range bound $\Delta\phi^e \leq 1$ on ekpyrotic models. This bound implies a minimum value of the parameter $\gamma^e,$ which for the cases of constant and scaling equation of state translates into the values
\begin{align}
& \gamma^e_{min,naive} = 1.9 \rightarrow \epsilon^e=7400\,, \\
& \gamma^e_{min,scaling} = 3.7 \rightarrow \epsilon^e_{60} = 1.2 \times 10^{7}\,.
\end{align}
As we can see, for the constant equation of state case the fast-roll parameter already has to be very high, and in the scaling case we must start out at ultra-fast-roll in order to stay within the specified field range. Hence here also the cosmological dynamics is driven to its extreme, and the field range bound is just as constraining for ekpyrotic models as it is for inflationary ones. In the scaling case, $60$ e-folds before the end of ekpyrosis the fast-roll parameter has to be more than a million times larger than the minimum value required for ekpyrosis. What consequences does this have for observational predictions? Here the situation in ekpyrosis is slightly different than that in inflation, as the ekpyrotic scalar itself does not lead to long-wavelength classical density fluctuations \cite{Battarra:2013cha}. Rather, ekpyrotic models produce structure by having a second field whose perturbations get enhanced. These entropy perturbations must subsequently be converted into curvature perturbations, typically after the end of the ekpyrotic phase. Thus the formation of structure is somewhat more dissociated from the background evolution -- nevertheless, we may demand equally simple scaling behaviour in this sector. 

There are two main classes of models: those with minimally coupled scalar fields \cite{Finelli:2001sr,Finelli:2002we,Lehners:2007ac} and those with non-minimal coupling \cite{Qiu:2013eoa,Li:2013hga,Ijjas:2014fja}. In the first instance, the scalar potential must be significantly curved in the field direction perpendicular to the ekpyrotic background trajectory, while the second class of models can operate with a flat transverse potential. For models with minimal coupling, the action contains a second scalar field $\chi$ and the total scalar potential is assumed to be of the form \cite{Buchbinder:2007tw,Lehners:2009ja}
\begin{align}
V(\phi,\chi) = V(\phi,0) \left[ 1 + \frac{\kappa_2}{2}\frac{V_{,\phi\phi}}{V(\phi,0)}\chi^2 + \frac{\kappa_3}{3!}\epsilon^{3/2}\chi^3 + \frac{\kappa_4}{4!}\epsilon^{2}\chi^4 + \cdots \right] \label{MinimalPot}
\end{align} 
where $\kappa_2 = \frac{\epsilon}{\epsilon_\chi}$ is the ratio between the equations of state associated with the two scalars, and which we will assume to be close to unity. The normalisation of the higher order $\kappa_{3,4}$ coefficients is such that one would expect them to be ${\cal O}(1)$ quantities in general \cite{Lehners:2013cka}. At the level of linear perturbation theory these models yield the observational predictions \cite{Lehners:2007ac,Buchbinder:2007tw,Ijjas:2013sua,Lehners:2015mra,Boyle:2003km,Baumann:2007zm}
\begin{align}
n_s  & = 1 - \frac{2}{3({\cal N}+1)^\gamma} - \frac{\gamma}{{\cal N}+1} +\frac{4}{3}(1-\kappa_2)\,, \\
\alpha_s & = - \frac{2\gamma}{3({\cal N}+1)^{\gamma+1}} - \frac{\gamma}{({\cal N}+1)^2} \,,\\
r & = r_{min}\,,
\end{align}
where we assumed that $\kappa_{2,{\cal N}}\approx 0.$ For the naive and the scaling equations of state, assuming exact scale freedom $\kappa_2=1,$ one thus obtains the respective predictions
\begin{align}
n^e_{s,naive} = 0.97 & \qquad n^e_{s,scaling} = 0.94 \,,\\
\alpha^e_{s,naive} = -5.0 \times 10^{-4} & \qquad \alpha^e_{s,scaling} = -1.0 \times 10^{-3}\,, \\
r^e_{naive} = r_{min} & \qquad r^e_{scaling} = r_{min}\,.
\end{align}
Again, we find that the spectral tilt is spot-on for the naive model, while the scaling model is redder, though not quite as red as for the inflationary scaling model. The running of the spectral index is significant, especially for the scaling model. Moreover, a general feature of ekpyrosis is that gravitational waves are not produced at linear order, since the universe slowly contracts during ekpyrosis, and gravitational waves are only sensitive to the evolution (here almost a non-evolution) of the scale factor. However, at second order in perturbation theory, the scalar fluctuations act as a source for gravitational waves, and this minimal signal, which however has frequency-dependent features \cite{Baumann:2007zm}, is the only primordial gravitational wave signal produced in these models -- this is why we state the tensor-to-scalar ratio as $r_{min}.$ (We note in passing that more involved models might generate gravitational waves due to couplings to additional fields, see e.g. \cite{Ben-Dayan:2016iks,Ito:2016fqp}.)

The models with non-minimal kinetic coupling assume an action of the form \cite{Qiu:2013eoa,Li:2013hga,Ijjas:2014fja}
\begin{align}
S = \frac{1}{2}\int d^4 x \sqrt{-g} \left[ R - (\partial \phi )^2 - \Omega^2(\phi) (\partial\chi)^2 - 2 V(\phi) \right]\,,
\end{align}
where the non-minimal coupling $\Omega(\phi)$ plays the role of an effective equation of state for the scalar $\chi,$ which remains frozen at background level during the ekpyrotic phase. At large ${\cal N},$ we may approximate the ekpyrotic potential by $V(\phi) \approx - V_0 e^{-\sqrt{2\epsilon}\phi}$ for a constant $V_0,$ and the coupling function by  $\Omega^2(\phi) \approx e^{-\sqrt{2\kappa_2\epsilon}\phi}.$ One then obtains the observational predictions \cite{Qiu:2013eoa,Li:2013hga,Fertig:2013kwa,Lehners:2015mra}
\begin{align}
n_s  & = 1 + 2 (1-\sqrt{\kappa_2}) - \frac{7\gamma}{3({\cal N}+1)}\,, \\
\alpha_s & = - \frac{7\gamma}{3({\cal N}+1)^2} \,,\\
r & = r_{min}\,.
\end{align}
For the naive and the scaling equations of state, assuming again $\kappa_2=1,$ one thus obtains the respective predictions 
\begin{align}
n^e_{s,naive} = 0.93 & \qquad n^e_{s,scaling} = 0.86 \,,\\
\alpha^e_{s,naive} = -1.2 \times 10^{-3} & \qquad \alpha^e_{s,scaling} = -2.3 \times 10^{-3}\,, \\
r^e_{naive} = r_{min} & \qquad r^e_{scaling} = r_{min}\,.
\end{align}
The scalar tilt comes out as very red, and with a significant running in both cases.

\section{Comparison to observations and small deformations}

We can now compare our models to current observational bounds in a little more detail. Measurements of the cosmic microwave background by the Planck satellite yield the following observed parameters (stated at one sigma accuracy, except for the upper bound on the tensor-to-scalar ratio) \cite{Ade:2015xua,Ade:2015lrj}:
\begin{align}
n_s & = 0.968 \pm 0.006 \,,\\
r & < 0.07 \quad(95\% \, C.L.) \,,\\
\alpha_s & = -0.003 \pm 0.007 \,.
\end{align}
The most obvious feature of the small-field scale-free models described above is that they {\it all} predict a tilt that is too red!  This discrepancy can however be remedied by allowing small deviations from exactly scale-free power law relations. For instance, for inflation we may consider models where the slow-roll parameter evolves according to \cite{Mukhanov:2013tua,Ijjas:2013sua}
\begin{align}
\epsilon^i = \frac{\delta}{(N+1)^{\gamma^i}}\,,
\end{align}
where we have allowed for an overall coefficient $\delta.$ This changes the field range to
\begin{align}
\Delta\phi^i = \frac{2\sqrt{2}}{\gamma - 2}\delta \left[ (N+1)^{\frac{2-\gamma}{\gamma}} - 1\right]\,,
\end{align}
while observational parameters are modified according to
\begin{align}
n_s  & = 1 - \frac{2\delta}{(N+1)^\gamma} - \frac{\gamma}{N+1}\,, \\
\alpha_s &  =  - \frac{2\gamma\delta}{(N+1)^{\gamma+1}} - \frac{\gamma}{(N+1)^2}\,, \\
r &= 16 \epsilon\,.
\end{align}
We can then lower the value of $\gamma^i$ in order to obtain a spectral index that is not quite as red, while simultaneously choosing $\delta$ small enough that the field range bound is still satisfied. The most favourable case is obtained by the combination $\gamma^i=2.4,\delta=0.25,$ which corresponds to $\epsilon^i_{60} = 1.3 \times 10^{-5}.$ This brings the tilt to within one sigma of the observed value and leads to the ``predictions''
\begin{align}
n_s = 0.96, \qquad \alpha_s = -6.5\times 10^{-4}, \qquad r=2.1 \times 10^{-4}\,.
\end{align}
The running is then again smaller, while the tensor-to-scalar ratio is increased, though still quite small in terms of expectations for a near future detection.

Similarly, in the minimally coupled ekpyrotic case we could consider
\begin{align}
\epsilon^e = 3 \delta ({\cal N}+1)^\gamma
\end{align}
This would modify the field range to
\begin{align}
\Delta\phi^e = - \sqrt{6\delta}\int \frac{({\cal N}+1)^{\gamma/2}}{3\delta({\cal N} +1)^\gamma-1} d{\cal N}\,,
\end{align}
while the observational parameters are changed to
\begin{align}
n_s  & = 1 - \frac{2}{3\delta({\cal N}+1)^\gamma} - \frac{\gamma}{{\cal N}+1}\,, \\
\alpha_s & = - \frac{2\gamma}{3\delta({\cal N}+1)^{\gamma+1}} - \frac{\gamma}{({\cal N}+1)^2}\,. 
\end{align}
Here the combination $\gamma^e=2.4, \delta =5.3$ is optimal, in that it leads to
\begin{align}
n_s = 0.96, \qquad \alpha_s = -6.5 \times 10^{-4}, \qquad \epsilon^e_{60} = 3.1 \times 10^5\,.
\end{align}

For ekpyrotic models with non-minimal kinetic coupling, we may exploit the freedom in the coupling function to obtain a spectral index in accord with observational bounds, while maintaining the exact scale-free evolution for the background. For instance, if we choose a constant ratio between the equations of state implied by the potential and coupling function, in particular $\kappa_2 = 0.90,$ we obtain
\begin{align}
n_s = 0.96, \qquad \alpha_s = -2.3\times 10^{-3}, \qquad \epsilon^e_{60}=1.2\times 10^7\,.
\end{align} 
In this case only the spectral index is changed, while the expectation remains for the running to be substantial. 

So far we have focussed on linear perturbations, but interesting additional constraints arise when considering corrections to a Gaussian distribution of perturbations, especially for ekpyrotic models. Current observational bounds on local non-Gaussianity are \cite{Ade:2015ava}
\begin{align}
f_{NL} & = 0.8 \pm 5.0\,,\\
g_{NL} & = (7.9 \pm 9.0) \times 10^4\,,
\end{align}
where $f_{NL}$ is the local bispectrum parameter and $g_{NL}$ the local trispectrum parameter, which govern the sizes of any quadratic and cubic corrections to the real space curvature fluctuation on uniform density hypersurfaces $\zeta$ \cite{Komatsu:2001rj},
\begin{align}
\zeta=\zeta_L + \frac{3}{5}\zeta_L^2 + \frac{9}{25} \zeta_L^3\,. \label{zeta}
\end{align} 
The inflationary models we were led to are all of an ultra-slow-roll type with very small values of the slow-roll parameter during the time when observable fluctuations are amplified. This immediately implies that these models have tiny self-interactions of the inflaton field, and lead to absolutely tiny corrections to Gaussianity easily in agreement with the observational bounds. By contrast, the ekpyrotic models are ultra-fast-roll, such that one may expect the opposite situation, i.e. one expects the deviations from Gaussianity to be substantial. Here again there is a big difference between the models with minimal coupling and those with a non-minimal kinetic coupling. The minimally coupled models have a scalar potential of the form \eqref{MinimalPot}, where $\kappa_{3,4}$ are order $1$ numbers that parameterise the skewness of the potential and its quartic correction around the ekpyrotic background trajectory. These higher order terms in the potential lead to non-Gaussian corrections to the entropy perturbation already during the ekpyrotic phase \cite{Lehners:2007wc}, and during conversion from entropy to curvature fluctuations these corrections get transferred to non-Gaussian features of the curvature perturbations \cite{Buchbinder:2007at,Lehners:2009qu}. By contrast, the non-minimally coupled models have a flat potential in the entropy direction (by assumption), and thus they do not generate non-Gaussianity during the ekpyrotic phase \cite{Fertig:2013kwa}. However, the conversion process itself is necessarily non-linear, and in and of itself it induces a certain amount of non-Gaussianity that was determined by numerical studies in \cite{Lehners:2008my,Fertig:2015ola}. The predictions for both types of models can be summarised as follows
\begin{align}
 f_{NL,minimal}^e \sim \kappa_3 \sqrt{\epsilon_{60}} \gtrsim \kappa_3 10^3 \qquad & f_{NL,kinetic}^e \sim \pm 5 \\
 g_{NL,minimal}^e \sim {\cal O}(1) \left(\frac{\kappa_3^2}{80}+ \frac{\kappa_4}{60}-\frac{19}{60}\right) \epsilon_{60} \gtrsim  10^6 \qquad & g_{NL,kinetic}^e \sim - {\cal O}(10^3)
\end{align}
For minimally coupled models the bispectrum parameter $f_{NL}$ comes out as too large by two orders of magnitude, unless $\kappa_3$ is finely tuned to be small. One may imagine that a symmetry might require the potential to be even in $\chi,$ and this would imply that $\kappa_3=0.$ However, even in such a case, the trispectrum parameter is still very large, of order $10^5$ or $10^6$ at least, because there is a term in $g_{NL}$ that depends purely on $\epsilon_{60}$ independently of the characteristics of the potential. This is at least an order of magnitude larger than the observational bound, and in fact this parameter is so large that according to \eqref{zeta} one may question the validity of perturbation theory. Hence we conclude that ultra-fast-roll minimally coupled models are in fact ruled out because of the large amount of non-Gaussianity that they produce. On the other hand, the models with non-minimal kinetic coupling have predictions that fit well with current observational bounds, though one would expect to observe a non-trivial bispectrum parameter $f_{NL}$ in upcoming CMB experiments (or, alternatively, one may try to identify a conversion mechanism that does not lead to large non-Gaussian corrections). Certainly, in view of these bounds, the kinetically coupled ekpyrotic models remain the most promising small-field models that can generate curvature perturbations in agreement with observations during a contracting phase.

\section{Discussion}

It is an interesting and highly welcome development that bounds that come from quantum gravity impose significant restrictions on cosmological models. By imposing the field range bound, i.e. the requirement that a scalar field should not change by more than one Planck mass during the course of its evolution (as the effective theory that one uses would otherwise lose its validity), combined with the physical requirement of demanding a scale-free power law evolution of the equation of state, one obtains severe restrictions on both inflationary and ekpyrotic models. In all cases the dynamics is driven to rather extreme parameter ranges, namely ultra-slow-roll for inflation and ultra-fast-roll for ekpyrosis. One striking consequence of these bounds is that all such models end up predicting a tilt of the scalar spectral index that is too red by many standard deviations to agree with observations. As we saw, one is thus forced to give up exact scale freedom. By allowing for small deviations from an exact scale-free power law, the spectral index can be brought into agreement with observations, but at the expense of diluting some of the physical motivation for these models. There are however other generic predictions, which are not significantly altered in the models deviating slightly from scale freedom: in particular the spectral running in all models is at a level of ${\cal O}(10^{-4})$ to ${\cal O}(10^{-3})$ and in all cases negative in sign, leading to a clear observational signature that one can look for in upcoming CMB experiments. Meanwhile, the level of predicted gravitational waves is very low, even in the inflationary case. This is certainly in agreement with the current upper bounds, but unfortunately means that for these models one would not expect a detection of primordial gravity waves any time soon. The biggest discrepancy between the inflationary and ekpyrotic models remains at the level of non-Gaussianity. Ultra-fast minimally coupled ekpyrotic models are in fact already ruled out by current observational upper bounds, while models with non-minimal kinetic coupling lead to interesting expectations regarding upcoming CMB observations.

We should note that the field range bound has further implications for ekpyrotic models. As we discussed above, in ekpyrotic models entropy perturbations are generated first, and these are then converted into curvature perturbations, either before or after a cosmological bounce \cite{Fertig:2016czu}. We applied the field range bound to the ekpyrotic phase alone. But the conversion phase will require its own region in moduli space, and similarly for the bounce.  In fact, in the models considered so far conversion happens during a phase of kinetic domination, during which the fields evolve according to $\phi(t)=-\ln(\pm \sqrt{\frac{3}{2}}t) + \phi_0,\quad a(t)=a_0 (-t)^{1/3}$ with $\phi_0,a_0$ being integration constants. This immediately implies that $\Delta\phi_{conversion} = \ln(\sqrt{\frac{2}{3}}t_{beg}/t_{end})$ where $t_{beg,end}$ denote the beginning and end of the conversion phase. If the conversion phase lasts for about one e-fold of evolution, implying $t_{beg}/t_{end}\approx e^{3/2},$ then the field range is $\Delta\phi_{conversion} \sim {\cal O}(1).$ Thus the conversion phase itself takes up about one Planck mass in terms of field range. The picture that emerges is then that there might be different effective theories for the separate phases of ekpyrosis, conversion and bounce. From a fundamental physics point of view this is certainly not unexpected, but it means that a precise description of ekpyrotic models in quantum gravity will entail an understanding of these different effective theories, and of the transitions between them. We may thus still be quite a long way away from a fundamental understanding of such models.

\section{Comments on the slope bound}

In closing, it is interesting to also discuss a recent bound on the slope of scalar potentials proposed in \cite{Obied:2018sgi,Agrawal:2018own} (and see also \cite{Andriot:2018wzk}), namely that at positive values of the potential the slope must be at least as large as the potential value itself,
\begin{align}
\frac{|\nabla V|}{V} \gtrsim {\cal O}(1)\,, \qquad (V>0)\,.
\end{align}
This means that for positive potentials one would expect $\epsilon \gtrsim {\cal O}(1).$ Such a bound is conjectured from many examples of string compactifications and, if true, would have very significant implications for cosmology. For example, it would immediately rule out all of the ultra-slow-roll models considered in the present paper! However, it would not restrict ekpyrotic models, as these operate at negative values of the potential. More generally, such a bound imposes strong restrictions on inflationary models, and it will be important to clarify what the precise value of the ${\cal O}(1)$ minimum slope really is. 

Such a bound would also have important ramifications for the global structure of spacetime, as it might render false-vacuum eternal inflation inoperative (as one could always roll out of a local ``minimum'' and thus the necessary time-scale for Coleman-De Luccia type tunneling \cite{Coleman:1980aw} to occur would simply not be available), while slow-roll eternal inflation would be pushed to extremely high values of the potential where one might not trust the theory anymore.

One may wonder how one can see (on a more technical level) where the difficulty in having a small slope at positive values of the potential could come from. The supergravity approximation to string theory offers some insights in this respect. In particular, in 4-dimensional models with minimal supersymmetry, the scalar fields are described by a K\"{a}hler potential $K(A,A^\star),$ which is a hermitian function of a complex scalar field $A,$ and a superpotential $W(A),$ which is a holomorphic function of $A.$ The kinetic term is then given by $-K_{,AA^\star} \partial^\mu A \partial_\mu A^\star,$ and the potential by the famous formula \cite{Stelle:1978ye,Ferrara:1978em,Cremmer:1978iv}
\begin{align}
V=e^K \left[ K^{,AA^\star} [(\partial_A + K_{,A}) W(A)][(\partial_{A^\star} + K_{,A^\star}) W^\star(A^\star)] - 3 W W^\star \right]\,. \label{sugrapot}
\end{align}
As a representative example, consider the K\"{a}hler potential $K=-\frac{1}{2}(A-A^\star)^2.$ If one writes the complex scalar in terms of two real scalars, $A =\frac{1}{\sqrt{2}}(\phi + i \chi),$ then this choice leads to a canonically normalised kinetic term
\begin{align}
-K_{,AA^\star} \partial^\mu A \partial_\mu A^\star=-\partial^\mu A \partial_\mu A^\star = -\frac{1}{2}(\partial\phi)^2 -\frac{1}{2}(\partial\chi)^2\,.
\end{align}
Meanwhile, with a simple choice for the superpotential $W = w_0 e^{cA},$ where $w_0, c$ are constants, one obtains the scalar potential (in the $\phi$ direction)
\begin{align}
V(\phi) = |w_0|^2 (c^2-3) e^{\sqrt{2}c \phi}\,.
\end{align}
This form of the potential is both typical and suggestive: one can obtain a positive potential for $c^2 > 3,$ which immediately implies that it will be too steep as it implies  $\epsilon > 3,$ or one can obtain a negative potential for $c^2 < 3,$ but this will then be too shallow for ekpyrosis as $\epsilon < 3$ (an ekpyrotic model embedded in supergravity has been constructed in \cite{Koehn:2013upa} -- this model uses a much more elaborate form for the superpotential and results in $\epsilon$ only slightly smaller than $3$ and consequently involves a very large field range). One thus typically obtains exactly the opposite of what would be required in both types of cosmological models. The formula \eqref{sugrapot} for the potential illustrates the general behaviour: large derivatives cause the first term in brackets to grow, thus raising the potential, while small derivatives render the second, negative term more important. The preference is clearly for a positive steep potential or a negative shallow one, and this is a source of tension for both inflationary and ekpyrotic models (we note that a no-go theorem to that effect was recently exhibited in the context of ten-dimensional type IIB supergravity in \cite{Uzawa:2018sal}). The goal then is to either find counterexamples to these slope bounds, or to start thinking about cosmological models that can fit all the bounds in a natural way and at the same time agree with observations of the sky.

\acknowledgments

I would like to thank Sebastian Bramberger, Alice Di Tucci, Paul Steinhardt, Kelly Stelle and Cumrun Vafa for valuable discussions.

\bibliographystyle{utphys}
\bibliography{SmallField}

\end{document}